\RequirePackage{fancyvrb}
\documentclass[submission,copyright,creativecommons]{eptcs}
\providecommand{\event}{ICLP 2026}

\usepackage{iftex}

\ifpdf
  \usepackage{underscore}         % Only needed if you use pdflatex.
  \usepackage[T1]{fontenc}        % Recommended with pdflatex
\else
  \usepackage{breakurl}           % Not needed if you use pdflatex only.
\fi

\usepackage{amsmath}
\usepackage{amssymb}
\usepackage{comment}
\usepackage{cprotect}
\usepackage{ebproof}
\usepackage{fancyvrb}
\usepackage{fvextra}
\usepackage{graphicx}
\usepackage{latexsym}
\usepackage{mathtools}
\usepackage{multirow}
\usepackage[all=normal, leading=tight]{savetrees}
\usepackage{simplebnf}
\usepackage{subcaption}
\usepackage{tikz}
\usepackage{wasysym}
\usepackage{xcolor}
\usetikzlibrary{positioning}

\fvset{commandchars=\\\{\}, fontsize=\small}

\newcommand{\ret}{\ensuremath{\hookrightarrow{}}}
\let\unlinkedfootref\footref
\renewcommand{\footref}[1]{\hyperlink{#1-target}{\NoHyper\unlinkedfootref{#1}\endNoHyper}}
\newcommand{\footreflabel}[1]{\hypertarget{#1-target}{}\label{#1}}

\newif\ifdraft
\draftfalse   % enable draft mode (todo comments)

\newcommand{\mk}{{\sc miniKanren}}
\newcommand{\ck}{{\sc chrKanren}}
\newcommand{\uk}{$\mu${\sc Kanren}}

\newcommand{\rs}[1]{\ifdraft{\color{red} [#1 - Raffi]}\fi}
\newcommand{\cm}[1][\relax]{\ifdraft{\color{blue}[CITE ME\ifx#1\relax]\else: #1]\fi}\fi}

\newcommand{\lc}[1]{$\lambda$-calculus}

\title{\ck: Constraint Handling Rules in a Relational Language}
\author{Rafaello Sanna
  \institute{Harvard University\\ Massachusetts, USA}
  \email{rsanna@g.harvard.edu}
  \and
  William E. Byrd
  \institute{University of Alabama at Birmingham\\ Alabama, USA}
  \email{webyrd@uab.edu}
  \and
  Nada Amin
  \institute{Harvard University\\ Massachusetts, USA}
  \email{namin@seas.harvard.edu}
}
\def\titlerunning{\ck}
\def\authorrunning{Sanna et al.}
\begin{document}
\maketitle

\VerbatimFootnotes

\begin{abstract}
We present \ck{}, a dialect of the purely relational constraint logic programming language \mk~\cite{reasoned,reasoned2,byrdthesis} which includes support for Constraint Handling Rules (CHR)~\cite{fruhwirth1995constraint}, a language for writing rule-based programs such as constraint solvers. We show how to integrate CHR's constraint propagation mechanism into the language of \mk{} search streams such that both processes remain complete. We also use \ck{} to illustrate novel applications of constraints in \mk{}, such as semantic unification of user-defined data structures and example propagation in relational interpreters in the style of \textsc{Myth}~\cite{Osera:2015:TPS:2737924.2738007}.
\end{abstract}

\ifdraft
\section*{Reviewer Notes}

\begin{enumerate}
  \item all rules seem to be single-headed (R1)
  \item Overall novelty seems somewhat limited. (R1)
  \item \checkmark \S 2.2: the general rule form seems to be missing the guard which only shows up first in the example (R1)
  \item \checkmark \S 2.2: the Figure ?? reference is broken (R1, R2)
  \item \S 4: Can you elaborate on why the double rules are needed \verb|(Xo sym)| / \verb|(ground X? sym)| (R1)
  \item \checkmark \S 5: Fig. 9(a) cases (ii) and (iv)  have typos (R1)
  \item Many code snippets/tables are minimally explained (R2)
  \item \checkmark Figure 5, 6, and 7 are not referenced (R2)
  \item The overall relevance and significance of the work are somewhat unclear (R2)
  \item \checkmark Paragraph 3: "variables x" -> in line 8 of Figure 1 there is no x (R2) \rs{I don't know what the reviewer meant here -- maybe just clarify that section in general?}
  \item \checkmark Paragraph 11: "part of [11]’s solver" -> "part of the solver proposed by Dovier et al. [11]" (R2)
  \item The main issues are relevance to ICLP (R3)
  \item The main issues are ..., potential impact of the contribution, which is limited to \mk{} users (R3)
  \item \checkmark I suggest the authors to read [FreeCHR] (R3)
\end{enumerate}
\fi

\maketitle

\section{Introduction}
\label{sec:intro}

\mk{} is a logic programming language focused on \textit{relational purity}, the ability to write programs which run completely in all groundedness modes~\cite{byrdthesis}. One way \mk{} realizes this goal is with built-in support for Constraint Logic Programming (CLP)~\cite{Jaffar1994CLPSurvey}, which lazily propagates constraints over logic variables to prune large swathes of the search tree. CLP is effective, but only for domains with bespoke solvers, which are hard or impossible to write with only relational primitives. We present \ck{}, which integrates Constraint Handling Rules (CHR) --- a declarative formalism for writing rule-based constraint solvers --- into \mk{}. \ck{} allows users to write their own constraint solvers without needing to modify or understand low-level internals.

\subsection{Motivating Example: Relational Interpretation with Constraints}
\label{sec:example}

A relational interpreter~\cite{byrd2017,byrdMiniKanrenLiveAndUntagged} is an interpreter written as a pure relation. When run under a complete search strategy, a relational interpreter can be used not only to evaluate ground programs, but also to synthesize programs from output values, complete partial programs, or generate quines. Consider, for example, \verb|evalo|\footnote{The suffix \verb|-o| or \verb|-|$^o$ is a convention used to denote that a host-language identifier refers to a relation. We define \verb|evalo| in Listing \ref{fig:evalo}.}: a relational interpreter for a \lc{} dialect with pairs and numbers\footnote{Because \ck{} is embedded in Scheme and the language we interpret is also Scheme-like, it's easy to confuse terms in the guest and host languages. To make this dichotomy clearer, we use the name ``\lc{}'' for the guest language and the name ``Scheme'' for the host language. \lc{} terms are denoted by Scheme S-expressions and constructed using Scheme's quasiquote (\verb|`|) and unquote (\verb|,|) syntax. For example, if the Scheme variable \verb|a| refers to the Scheme S-expression \verb|(x 3)|, then the Scheme expression \verb|`(lambda (x) ,a)| would evaluate to the Scheme S-expression \verb|(lambda (x) (x 3))| and denote the \lc{} term $\lambda x . x\ 3$.}\rs{This footnote is colossal -- should we shorten it somehow?}. It relates a $\lambda$-term and an environment to its weak call-by-value normal form\rs{It seems like this language is kosher. See: \url{https://studwww.itu.dk/~sestoft/papers/sestoft-lamreduce.pdf}.}. We ask the system to find a \lc{} term which evaluates to a function such that when it is called in the empty environment, it evaluates to a pair whose first component is \verb|1|:

\begin{Verbatim}
(run 1 (p) (fresh (v q) (evalo `(,p ,v) '() `(1 . ,q))))
\ret ((((lambda (_.0) (cons 1 _.1))) (num _.1) (sym _.0)))
\end{Verbatim}

\noindent
The system returns one general class of solutions: those where
\verb|p| is a Scheme expression of the pattern \verb|(lambda (_.0) (cons 1 _.1))|, where \verb|_.0| and \verb|_.1| are fresh logic variables which must be a number and a symbol, respectively. This Scheme expression denotes a $\lambda$-term which ignores its argument, then constructs the requested pair, assuming the second \verb|cons| element to be a number literal.

While this interpreter can uniformly solve a number of programming problems~\cite{byrd2017}, it can be unsatisfying when solving other queries. Consider the following query, which asks for a \lc{} term \verb|q|, such that when evaluated as the first argument of a \verb|cons| call with \verb|2| as its second argument, it evaluates to the pair \verb|(1 . 3)|:

\begin{Verbatim}
(run 1 (q) (evalo `(cons ,q 2) '() '(1 . 3)))
;; runs forever...
\end{Verbatim}

\noindent
It's clear that no such term exists, as \verb|2| cannot evaluate to \verb|3|. This causes \verb|evalo| to diverge, which is a sound response under \mk{}'s semi-decidable semantics, though not an ideal one. We would prefer that \verb|evalo| exhibit finite failure in this case: to prove that there are no valid instantiations of \verb|q|.

Constraints allow us to do exactly this: We can, using simple CHR rules, write a constraint-based interpreter \verb|eval^o| which decomposes the same query into independent sub-constraints over each component of the pair. Because constraint propagation is driven by the immediate availability of information, rather than traditional search, the sub-constraint over \verb|2| fails immediately:

\begin{Verbatim}
(run 1 (q) (eval^o `(cons ,q 2) `() `(1 . 3)))
\ret ()
\end{Verbatim}

\ck{} allows users to mix traditional \mk{} search with user-defined constraint solvers such as \verb|eval^o|. The propagation behavior seen in this example is the natural behavior of CHR rules, which fire when arguments become sufficiently ground or restricted by other constraints. The constraint-based evaluator is just one example of what this enables. More generally, any domain in which constraints can propagate information during search benefits from the same approach.

\subsection{Contributions}

We make the following contributions:

\begin{enumerate}

  \item We augment the stream-based operational semantics of \mk{} with native support for CHR-style constraint propagation such that search and constraint propagation remain complete (Section~\ref{sec:semantics}).

  \item We implement the above semantics in a new \mk{} interpreter called \ck{} based on the first-order architecture of Rosenblatt et al.~\cite{Rosenblatt2019FirstOrderMK}.

  \item We demonstrate two applications uniquely enabled by the combination of \mk{}'s complete search and CHR's user-extensible constraint propagation: type-and-example-directed program synthesis (Section~\ref{sec:ex:synth}), and semantic unification over user-defined data structures (Section~\ref{sec:ex:sets}).
\end{enumerate}

\section{Preliminaries}

\subsection{\mk{}}
\label{sec:mk-intro}

\begin{figure}[h!]
\renewcommand{\figurename}{Listing}
\centering
\begin{subfigure}[t]{0.45\linewidth}
\begin{Verbatim}[numbers=left, fontsize=\scriptsize]
(define-relation (evalo exp env val)

  (conde

    [(numbero exp)
     (== val exp)]




    [(fresh (nm body)
       (== exp `(lambda (,nm) ,body))
       (symbolo nm)
       (== val `(clos ,nm ,body ,env)))]


    [(fresh (op arg)
       (== exp `(,op ,arg))
       (fresh (env^ nm body argv)
         (symbolo nm)
         (evalo op  env `(clos ,nm ,body ,env^))
         (evalo arg env argv)
         (evalo body `((,nm . ,argv) . ,env^) val)))]


    [(symbolo exp)
     (lookupo exp env val)]




    [(fresh (lhs rhs)
       (== exp `(cons ,lhs ,rhs))
       (fresh (lval rval)
         (== val (cons lval rval))
         (evalo lhs env lval)
         (evalo rhs env rval)))]))

\end{Verbatim}

\cprotect\caption{The \verb|evalo| relational interpreter.}
\label{fig:evalo}
\end{subfigure}
\hspace{1.3em}
\begin{subfigure}[t]{0.45\linewidth}
\begin{Verbatim}[fontsize=\scriptsize]
(define-constraint (eval^o exp env val))


(define-rules
  (forall (exp env val)
    (forget (eval^o exp env val))
    (ground number? exp)\footnote{\footreflabel{footnote:groundedness}This is not directly analogous to \mk{} type constraints in all groundedness modes, but we abbreviate for space. We expand on this further in Section \ref{par:higherlevel}.}
    =>
    (== exp val))

  (forall (env val nm body)
    (forget (eval^o `(lambda (,nm) ,body) env val))
    =>
    (symbolo nm)
    (== val `(clos ,nm ,body ,env)))

  (forall (op arg env val)
    (forget (eval^o `(,op ,arg) env val))
    =>
    (fresh (env^ nm body argv)
      (symbolo nm)
      (eval^o op env `(clos ,nm ,body ,env^))
      (eval^o arg  env argv)
      (eval^o body `((,nm . ,argv) . ,env^) val)))

  (forall (exp env val)
    (forget (eval^o exp env val))
    (ground symbol? exp)\footref{footnote:groundedness}
    =>
    (lookupo exp env val))

  (forall (lhs rhs env val)
    (forget (eval^o `(cons ,lhs ,rhs) env val))
    =>
    (fresh (lval rval)
      (== val (cons lval rval))
      (eval^o lhs env lval)
      (eval^o rhs env rval))))
\end{Verbatim}
\cprotect\caption{The \verb|eval^o| constraint interpreter.}
\label{fig:evalho}
\end{subfigure}
\cprotect\caption{Two implementations of an interpreter for the \lc{}. The \verb|lookupo| relation is user-defined.}
\label{fig:interpreters}
\end{figure}

We introduce \mk{} by defining the relational interpreter \verb|evalo|, as demonstrated in Section~\ref{sec:example}. Its complete definition is given in Listing~\ref{fig:evalo}. The interpreter is defined using the \verb|conde| goal constructor (line 3), which represents a disjunction of conjuncts. The first disjunct (line 5) handles numeric literals: the type constraint \verb|numbero| ensures the $\lambda$-term must be a Scheme number, and \verb|==| unifies the value with the term. Type constraints are disjoint: posting two conflicting type constraints to the same variable causes immediate failure without grounding the variable. The second disjunct (line 11) handles lambda abstractions: \verb|fresh| introduces new logic variables \verb|nm| and \verb|body|, which are unified against the structure of the term, and the value is unified with a closure object capturing the body and the current environment. The third disjunct (line 17) handles applications: the operator is evaluated to a closure and the operand to an argument by recursive calls to \verb|evalo|, and the closure body is then evaluated in the environment extended with the argument. The fourth disjunct (line 26) handles variable lookup. It uses \verb|symbolo|, a type constraint analogous to \verb|numbero|, which ensures the term is a symbol. The relation \verb|lookupo| retrieves the corresponding value from the environment. The fifth disjunct (line 32) handles the pair constructor \verb|cons|: the term is matched against a \verb|cons| form, and the value is required to be a pair whose components are the evaluations of the respective subterms.

\subsection{Constraint Handling Rules}
\label{sec:chr-intro}

Constraint Handling Rules (CHR) is a rule-based formalism for writing
constraint solvers~\cite{Fruhwirth1998CHR}. A CHR program maintains a multiset of constraints representing facts known to be true (the ``constraint store'') and a set of rules which manipulate it. Each rule fires when a matching combination of constraints is present in the constraint store, removing some constraints and adding new ones. This process continues until a fixpoint is reached. Unlike the search of \mk{}, CHR uses ``don't-care'' nondeterminism: when multiple rules are applicable, the choice of which to fire is arbitrary, and it is up to the rule author to ensure that the rules are confluent.

Rules in \ck{} have the form \verb|(forall (|$i$ ...\verb|)| $ch$ ... \verb|=>| $g$ ...\verb|)|, which can be read as ``if, for any terms $i$ ..., each $ch$ holds, then each $g$ must also hold''. Each $ch$ is a \textit{rule head}, which is either a constraint guard (written as a constraint, optionally annotated with the witness modifiers \verb|keep| or \verb|forget|), a syntactic unification guard (written with operator \verb|===|) or a predicate guard (written with operators \verb|ground| or \verb|scheme|). Constraint guards ensure certain constraints are currently in the constraint store, syntactic unification guards ensure that given terms unify, and predicate guards ensure that terms pass various host-language checks. The body $g$ ... is a conjunction of \mk{} goals which are executed when the rule fires. Additionally, any constraint guard marked with \verb|forget| is removed from the constraint store when the rule fires.

To illustrate these ideas, consider the constraint-based interpreter
\verb|eval^o| in Listing~\ref{fig:evalho}, which implements the
same interpreter as \verb|evalo| using CHR rules. Each rule corresponds directly, and is vertically aligned with, a corresponding disjunct in the original interpreter. The first rule (line 5) handles the evaluation of numbers: the \verb|(ground number? exp)| guard checks that \verb|exp| is a ground number, the evaluation constraint is removed from the constraint store, and the unification \verb|(== exp val)| goal is posed. The second rule (line 11) handles lambda abstractions: the term is matched against a \verb|lambda| form in the rule head, and the value is unified with a closure object. The third rule (line 17) handles applications: three sub-constraints are posted to evaluate the operator, operand, and body respectively. The fourth rule (line 26) handles ground symbol lookup, using the same \verb|lookupo| helper relation as in the corresponding case of Listing~\ref{fig:evalo}. The fifth rule (line 32) handles \verb|cons|: the constraint is decomposed into two sub-constraints over the components of the pair.

Figure~\ref{fig:trace} shows an execution trace of \verb|eval^o| on the query from Section~\ref{sec:example}. The initial constraint store contains a single constraint, \verb|(eval^o (cons q 2) () (1 . 3))|. Rule 5 (line 32) fires, decomposing the constraint into two sub-constraints: \verb|(eval^o q () 1)| and \verb|(eval^o 2 () 3)|. Rule 1 (line 5) then fires on the second sub-constraint, as \verb|2| is a ground number, reducing it to the unification \verb|2| $\equiv$ \verb|3|. This unification fails immediately, causing the entire query to fail finitely without considering the value of \verb|q|.

\SaveVerb{traceInit}|(eval^o (cons q 2) () (1 . 3))|
\SaveVerb{traceEvalQ}|(eval^o q () 1)|
\SaveVerb{traceEvalTwo}|(eval^o 2 () 3)|
\SaveVerb{traceEq}|(== 2 3)|
\SaveVerb{traceFail}|fail|

\begin{figure}[h!]
\begin{align}
  \{ \text{\UseVerb{traceInit}} \}
  &\mapsto{}_{\text{(Rule 5)}}\ \{
    \text{\UseVerb{traceEvalQ}},\
    \text{\UseVerb{traceEvalTwo}}
  \} \\
  &\mapsto{}_{\text{(Rule 1)}}\ \{
    \text{\UseVerb{traceEvalQ}},\
    \text{\UseVerb{traceEq}}
  \} \\
  &\mapsto{}_{\text{(Unify)}}\ \text{\UseVerb{traceFail}}
\end{align}
\cprotect\caption{A trace of \verb|eval^o| on the query
\verb|(run 1 (q) (eval^o `(cons ,q 2) `() `(1 . 3)))|,
showing finite failure without grounding \texttt{q}.}
\label{fig:trace}
\end{figure}

\section{The \ck{} Language}
\label{sec:chrkanrenref}

In order to give a formal account of \ck{}, we present both the syntax of the language in terms of a BNF grammar (Section \ref{sec:syntax})\footnote{The full \ck{} language supports some shorthand which is not reflected in the syntax or semantics. Namely, the \verb|fresh| goal constructor explicitly conjoins its body into a single goal with \verb|conj|, the \verb|conde| goal constructor is rewritten to a \verb|disj| of \verb|conj|'s, the \verb|conj| and \verb|disj| constructors are restricted to binary versions, constraint guards without a witness modifier are implicitly wrapped in a \verb|keep| modifier, and the \verb|ground| constraint guard is elaborated to the more primitive \verb|scheme| guard.} and a stream-based operational semantics (Section \ref{sec:semantics}).

\subsection{\ck{} Syntax}
\label{sec:syntax}

\begin{figure}[h]
\begin{bnf}
  $rn, cn, i$ : Host Identifiers :in: $\mathcal{I}$ ;;
  $t, pr$ : Terms :in: $\mathcal{T}$ ;;
  $m$ : Witness Modifier :in: \{ \texttt{forget}, \texttt{keep} \} ;;
  $d$ : Definitions ::=
  | \texttt{(define-relation (}$rn$ $i$ ...\texttt{)}
  $g$ ...\texttt{)} : relation definition
  | \texttt{(define-constraint (}$cn$ $i$ ...\texttt{)}\texttt{)} : constraint definition
  | \texttt{(define-rules} $r$ ...\texttt{)} : rule definition
  ;;
  $r$ : Rules ::= \texttt{(forall (}$i$ ...\texttt{)} $ch$ ... \texttt{=>} $g$ ...\texttt{)} : rule syntax
  ;;
  $c$ : Constraints ::=
  | \texttt{(===} $t$ $t$\texttt{)} : syntactic unification
  | \texttt{(}$cn$ $t$ ...\texttt{)} : user constraint
  ;;
  $w$ : Witness ::=
  | \texttt{(}$m$ ($cn$ $t$ ...)\texttt{)} : witness
  ;;
  $ch$ : Rule Heads ::=
  | $w$ : constraint guard
  | \texttt{(scheme} $pr$ $t$ ...\texttt{)} : predicate guard
  | \texttt{(===} $t$ $t$\texttt{)} : syntactic unification guard
  ;;
  $g$ : Goals ::=
  | \texttt{succeed} : trivial goal
  | \texttt{fail} : impossible goal
  | \texttt{(conj} $g_1$ $g_2$\texttt{)} : goal conjunction
  | \texttt{(disj} $g_1$ $g_2$\texttt{)} : goal disjunction
  | \texttt{(fresh (}$i$ ...\texttt{)} $g$\texttt{)} : existential goal
  | \texttt{(}$rn$ $t$ ...\texttt{)} : relation call
  | $c$ : constraint call
  ;;
\end{bnf}
\caption{A grammar outlining the syntax of \ck{}.}
\label{fig:grammar}
\end{figure}

The primitive syntax of \ck{} is based on that of \uk{}~\cite{microKanren} and is outlined in Figure \ref{fig:grammar}. \ck{} is embedded in Scheme and so inherits its Lisp-style syntax. One way in which \ck{} differs from traditional \mk{} syntax is its inclusion of \texttt{===} in place of \texttt{==} in the core calculus of the language. \ck{}'s \texttt{===} relation denotes \textit{syntactic unification}, while \texttt{==} denotes \textit{semantic unification}, a user-extensible notion of unification further explained in Section \ref{sec:ex:sets}. The latter is provided in userspace, and is therefore not part of our syntax or semantics as a primitive.

Another point of note is that \ck{} allows arbitrary goals to occur in the body of a rule. Because this includes disjunctions, this makes \ck{} an implementation of CHR$^{\lor}$, or CHR extended to include both ``don't-know'' and ``don't-care'' nondeterminism.

\subsection{\ck{} Semantics}
\label{sec:semantics}

\newcommand{\bind}{\mathbin{\scriptstyle>\mkern-5mu\scriptstyle>\mkern-0.5mu\scriptstyle=}}
\newcommand{\looparrow}{\circlearrowright}
\newcommand{\nil}{[\:]}
\newcommand{\pausedprop}{\circledcirc{}\ }

\begin{figure}[h]
\begin{bnf}
  $\Gamma$ : Search State ::= $\langle \sigma , \theta , \omega \rangle$ ;;
  $s$ : Search Streams ::=
  | $\nil{}$          : failure stream
  | $\Gamma \dblcolon s$ : solution stream
  | $s_1 \oplus s_2$    : disjunction stream
  | $s \bind g$        : binding stream
  | $\Gamma \bowtie g$ : paused binding stream
  | $\looparrow{} s$ : propagating stream
  | $\pausedprop{} \Gamma$ : paused propagating stream
  ;;
\end{bnf}
\caption{A grammar outlining the structure of \ck{} stream objects}
\label{fig:streams}
\end{figure}

% Macro-defining-macro creating a new metafunction
\newcommand{\newitmf}[2]{\expandafter\newcommand\csname #1\endcsname[1]{\textit{#2}(##1)}}

\newitmf{start}{start}
\newitmf{step}{step}
\newitmf{propagate}{propagate}
\newitmf{inhabit}{match}
\newitmf{apply}{apply}
\newitmf{call}{call}
\newitmf{unify}{unify}
\newitmf{var}{var}
\newitmf{head}{head}

The semantics and implementation of \mk{} is traditionally phrased in terms of states: objects which reify the values held by variables and the constraints over those variables. In our semantics, states (written $\Gamma$) hold a variable substitution (written $\sigma$), a constraint store (written $\theta$), and a rule application history (written $\omega$). We write $\sigma[\textbf{fresh}\ i\ ...]$ to mean the substitution $\sigma$ extended with the identifiers $i\ ...$ bound to fresh logic variables.

The search process of \mk{} works by threading states through lazy, potentially infinite streams. Each goal might be thought of as a function from some initial state to a stream of all possible future states which satisfy the goal in question. Following Rosenblatt~\cite{Rosenblatt2019FirstOrderMK}, we restrict ourselves to a fixed set of primitive stream constructors, as shown in Figure \ref{fig:streams}. The first two ``mature'' stream constructors represent finite failure (written $\nil$) and a successful search result (written $\Gamma :: s$). The remaining ``immature'' stream constructors represent intermediate states of the search which require more work to complete.

\begin{table}[h]
  \small
  \begin{tabular}{|c|p{33em}|}
    \textbf{Name} & \textbf{Description} \\
    $\mathcal{I}$ & The set of all host-language identifiers \\
    $\mathcal{T}$ & The set of all host-language terms \\
    $\mathcal{V}$ & The set of all logic variables \\
    $\mathcal{R}$ & The set of all defined constraint-handling rules \\
    $\call{rn, t ...}$ & Instantiates the body of the relation named $rn$ with arguments $t$ ... \\
    $\unify{\sigma, t_1, t_2}_{v}$ & Returns the substitution which syntactically unifies $t_1$ and $t_2$ with variables in the set $v$, or $\bot$ if none exists \\
    $\apply{pr, t ...}$ & Applies a host-language predicate $pr$ to the sequence of host-language terms $t$ ...; returns $\top$ or $\bot$ based on whether the predicate holds. \\
    $\head{r}$ & Returns the set of identifiers bound by the head of the rule $r$ \\
  \end{tabular}
  \caption{Names and descriptions of metafunctions and objects}
  \label{fig:helpers}
\end{table}

\begin{figure}[h!]
  \begin{align*}
  \start{\mathtt{succeed}, \Gamma}          &= \Gamma :: \nil \\
  \start{\mathtt{fail}, \Gamma}             &= \nil \\
  \start{(\mathtt{conj}\ g_1\ g_2), \Gamma} &= (\Gamma \bowtie g_1) \bind g_2 \\
  \start{(\mathtt{disj}\ g_1\ g_2), \Gamma} &= (\Gamma \bowtie g_1) \oplus (\Gamma \bowtie g_2) \\
  \start{(\mathtt{fresh}\ (i\ ...)\ g), \langle \sigma, \theta , \omega \rangle} &= \start{g, \langle \sigma[\textbf{fresh}\ i\ ...] , \theta , \omega \rangle} \\
  \start{(rn\ t\ ...), \Gamma}            &= \Gamma \bowtie \call{rn, t\ ...} \\
  \start{(\texttt{===}\ t_1\ t_2), \langle \sigma, \theta , \omega \rangle} &= \begin{cases}
    \propagate{\langle  \sigma' , \theta , \omega \rangle} & \unify{\sigma, t_1, t_2}_{\mathcal{V}} = \sigma' \\
    \nil & \unify{\sigma, t_1, t_2}_{\mathcal{V}} = \bot \\
  \end{cases} \\
  \start{(cn\ t\ ...), \langle \sigma, \theta , \omega \rangle} &= \pausedprop \langle  \sigma , \theta \cup \{ (cn\ t\ ...) \} , \omega \rangle \\
  \\
  \step{\nil} &= \nil \\
  \step{\Gamma :: s} &= \Gamma :: s \\
  \step{s_1 \oplus s_2} &= \begin{cases}
    s_2 & \step{s_1} = \nil \\
    \Gamma :: (s_2 \oplus s_3) & \step{s_1} = \Gamma :: s_3 \\
    s_2 \oplus \step{s_1} & \text{otherwise} \\
  \end{cases} \\
  \step{s \bind g} &= \begin{cases}
    \nil & \step{s} = \nil \\
    (\Gamma \bowtie g) \oplus (s_1 \bind g) & \step{s} = \Gamma :: s_1 \\
    \step{s} \bind{} g & \text{otherwise} \\
  \end{cases} \\
  \step{\looparrow s} &= \begin{cases}
    \nil & \step{s} = \nil \\
    (\pausedprop \Gamma) \oplus (\looparrow s_1) & \step{s} = \Gamma :: s_1 \\
    \looparrow \step{s} & \text{otherwise} \\
  \end{cases} \\
  \step{\Gamma \bowtie g} &= \start{g, \Gamma} \\
  \step{\pausedprop \Gamma} &= \propagate{\Gamma} \\
  \end{align*}
  \caption{An operational semantics detailing how goals are converted to streams, and how streams are advanced in terms of the metafunctions and objects of Table~\ref{fig:helpers} and the constraint propagation mechanism of Figure~\ref{fig:rules}.}
  \label{fig:search}
\end{figure}

\begin{figure}[h]
\begin{align*}
  &\propagate{\langle  \sigma, \theta , \omega \rangle} = \begin{cases}
    \looparrow \start{g, \langle  \sigma, \theta - \{ j\ |\ (\texttt{forget} \ j) \in \{w\ ...\} \} , \omega \cup \{ \langle  r , w\ ... \rangle \} \rangle} \\ \hspace{1.5em} \exists r \in \mathcal{R}.\ \inhabit{r, \langle \sigma[\textbf{fresh}\ head(r)], \theta, \omega \rangle}_\epsilon = \langle g , w\ ... \rangle \land \langle  r , w\ ... \rangle \notin \omega \\
    \langle \sigma , \theta , \omega \rangle :: \nil \\ \hspace{1.5em} \forall r \in \mathcal{R}.\ \inhabit{r, \langle  \sigma[\textbf{fresh}\ head(r)], \theta, \omega \rangle}_\epsilon = \bot \\
  \end{cases} \\
  \\
  &\inhabit{(\texttt{forall}\ (i\ ...)\ \texttt{=>}\ g\ ...), \Gamma}_{w ...} = \langle (\texttt{conj}\ g\ ...) , w\ ... \rangle \\
  &\inhabit{(\texttt{forall}\ (i\ ...)\ (\mathtt{scheme}\ pr\ t\ ...)\ ch\ ...\ \texttt{=>}\ g\ ...), \langle \sigma, \theta , \omega \rangle}_{w ...} = \\
  &\hspace{1em}\begin{cases}
    \bot & \apply{pr, \sigma(t\ ...)} = \bot \\
    \inhabit{(\texttt{forall}\ (i\ ...)\ ch\ ...\ \texttt{=>}\ g\ ...), \langle \sigma, \theta , \omega \rangle}_{w\ ...} & \text{otherwise} \\
  \end{cases} \\
  &\inhabit{(\texttt{forall}\ (i\ ...)\ (\texttt{===}\ t_1\ t_2)\ ch\ ...\ \texttt{=>}\ g\ ...), \langle \sigma, \theta , \omega \rangle}_{w ...} = \\
  &\hspace{1em}\begin{cases}
    \inhabit{(\texttt{forall}\ (i\ ...)\ ch\ ...\ \texttt{=>}\ g\ ...), \langle \sigma', \theta , \omega \rangle}_{w...} & \unify{\sigma, t_1, t_2}_{i...} = \sigma' \\
    \bot & \unify{\sigma, t_1, t_2}_{i...} = \bot \\
  \end{cases} \\
  &\inhabit{
      (\texttt{forall}\ (i\ ...)\ (m\ (cn\ t\ ...))\ ch\ ...\ \texttt{=>}\ g\ ...),
      \langle \sigma, \theta , \omega \rangle
  }_{w...} = \\
  &\hspace{1em}\begin{cases}
    \inhabit{
        (\texttt{forall}\ (i\ ...)\ ch\ ...\ \texttt{=>}\ g\ ...),
        \langle \sigma', \theta , \omega \rangle
    }_{(m\ (cn\ t'\ ...)), w\ ...} \\
    \hspace{3em} \exists\ (cn\ t'\ ...) \in \theta - \{ j\ |\ \exists m'.\ (m'\ j) \in \{w \ ...\} \} . \unify{\sigma, (t\ ...), (t'\ ...)}_{i...} = \sigma' \\
    \bot \\
    \hspace{3em} \forall\ (cn\ t'\ ...) \in \theta - \{ j\ |\ \exists m'.\ (m '\ j ) \in \{ w\ ... \} \} . \unify{\sigma, (t\ ...), (t'\ ...)}_{i\ ...} = \bot \\
  \end{cases} \\
\end{align*}
\caption{An operational semantics detailing the process of rule search and matching.}
\label{fig:rules}
\end{figure}

The disjunction constructor (written $s_1 \oplus s_2$) represents a stream whose elements are drawn from either of its two arguments and is used to implement disjunction. The bind constructor (written $s \bind g$) applies the goal of its right argument to each of the results from the stream on the left. Like the monadic bind after which it was named~\cite{Moggi1991Monads,Wadler1992Essence}, it forces a left-to-right evaluation order. The paused binding constructor (written $\Gamma \bowtie g$) delays the execution of a particular goal against a state. Paused binding constructors are implicitly added to conjunctions, disjunctions and calls to relations in order to support recursive relations and commutative disjunction. The two constraint propagation constructors (written $\looparrow s$ and $\pausedprop \Gamma$) are unique to \ck{}: The propagating constructor ($\looparrow s$) propagates constraints over each state element of a stream. The paused propagating constructor ($\pausedprop \Gamma$) delays constraint propagation against a state in a manner analogous to the paused binding constructor.

Goals are translated to streams with the $\start{g, \Gamma}$ metafunction, then advanced in small steps by the $\step{s}$ metafunction, as shown in Figure \ref{fig:search}. Both of these metafunctions make use of the $\propagate{\Gamma}$ metafunction, which actually performs constraint propagation.

The $\propagate{\Gamma}$ metafunction, as defined in Figure ~\ref{fig:rules}, takes a state $\Gamma$ and propagates its existing constraints. It does so by first finding a rule which can be applied to the state. If one exists, the rule is matched against the state, and the body of the rule is applied. If no such match is found, then the propagation is finished. The rule matching itself is performed by the $\inhabit{r, \Gamma}_{w ...}$ metafunction, which checks if a rule $r$ can match against the current state $\Gamma$. The index $w\ ...$ is used as an accumulator of ``witnesses'', annotated constraints we have already matched against that should not be considered for further head matches. The witnesses serve two roles: they prevent a single rule head from matching the same stored constraint twice, and they record which matched constraints should be removed when a rule fires. The matching against clause-heads occurs left-to-right and fails if any clause or guard should fail to find a match. Once this is done, \textit{match} returns both the goal of the matched rule and the witnesses the rule matched against.

\section{Application: Type-and-Example-Directed Program Synthesis}
\label{sec:ex:synth}

Section~\ref{sec:example} showed that \ck{} can improve traditional relational interpreter synthesis by evaluating subterms only as information becomes available. However, types have long been used in program synthesis as a means of both specifying the desired program and pruning the search space~\cite{Osera:2015:TPS:2737924.2738007}. Despite this, a relational type-checker in \mk{} cannot serve this role directly: conjoining a typing relation with a relational interpreter produces generate-and-test behavior, in which the interpreter generates candidate programs and the type-checker discards the ill-typed ones. Writing the type system as a \ck{} constraint resolves this: because constraints propagate incrementally during search rather than running as a separate phase, \verb|type^o| prunes ill-typed candidates as they are constructed without first enumerating them.

We demonstrate this through a CHR type system for the simply-typed \lc{}. The constraint \verb|type^o|, defined in Listing~\ref{fig:typeo}, relates an environment to a $\lambda$-term and a type. To illustrate the interaction between \verb|type^o| and \verb|evalo|, consider $\Omega$ --- the canonical diverging term of the untyped \lc{}, which is untypeable in the simply-typed \lc{}:

\begin{figure}[h!]
\renewcommand{\figurename}{Listing}
\begin{Verbatim}[fontsize=\footnotesize]
(define-constraint (type^o env exp typ))

(define-rules
  (forall (env n typ)
    (forget (type^o env n typ)) (numbero n) => (== typ `number))
  (forall (env n typ)
    (forget (type^o env n typ)) (ground number? n) => (== typ `number))
  (forall (env sym typ)
    (forget (type^o env sym typ)) (symbolo sym) => (lookupo sym env typ))
  (forall (env sym typ)
    (forget (type^o env sym typ)) (ground symbol? sym) => (lookupo sym env typ))
  (forall (env rator arglist body typ rst)
    (forget (type^o env `(,rator ,arglist ,body . ,rst) typ))
    =>
    (== rst `())
    (fresh (x i o)
      (== rator `lambda)
      (== arglist (list x))
      (symbolo x)
      (== typ `(-> ,i ,o))
      (type^o (cons (cons x i) env) body o)))
  (forall (env rator rand typ)
    (forget (type^o env `(,rator ,rand) typ))
    =>
    (fresh (t_input)
      (type^o env rator `(-> ,t_input ,typ))
      (type^o env rand t_input))))
\end{Verbatim}
\caption{A CHR type system for the simply-typed \lc{}.}
\label{fig:typeo}
\end{figure}

\begin{Verbatim}
(define omega `((lambda (x) (x x)) (lambda (x) (x x))))
\end{Verbatim}

\noindent Asking \verb|evalo| to evaluate $\Omega$ diverges, as
expected:

\begin{Verbatim}
(run 1 (p) (evalo omega `()  p))
;; runs forever...
\end{Verbatim}

\noindent By posting \verb|type^o| as a constraint before
evaluation, \ck{} proves $\Omega$ untypeable and fails finitely,
without ever invoking the evaluator:

\begin{Verbatim}
(run 1 (p) (fresh (t) (type^o `() omega t))
           (evalo omega `() p))
\ret ()
\end{Verbatim}

\section{Application: Semantic Unification}
\label{sec:ex:sets}

One of the primary advantages of CHR is its extensibility: new behavior can be introduced by adding rules over existing constraints without modifying the underlying solver or language runtime. \ck{} takes advantage of this by expressing unification itself as a collection of CHR constraints and rules. By reifying unification as a CHR constraint, \ck{} allows users to extend unification for new data structures in a principled and declarative manner.

To support this design, \ck{} distinguishes between two notions of unification. The first is \emph{syntactic unification}, written using the constraint \verb|===|. This form of unification is used internally when matching rule heads during CHR propagation. In contrast, the familiar \verb|==| relation of \mk{} is reinterpreted in \ck{} as \emph{semantic unification}: a user-extensible constraint whose meaning is determined entirely by CHR rules. This separation allows the operational semantics of CHR to remain stable, while exposing unification as an object of user-level definition.

Semantic unification over data structures is done using traditional CHR rules, using syntactic unification to bind concrete structure as appropriate. For example, traditional structural unification over cons cells is implemented by the following rule, which decomposes equality of compound terms into equality of their components:

\begin{verbatim}
(forall (a b c)
  (forget (== (cons a b) c))
  =>
  (fresh (d e) (=== c (cons d e)) (== d a) (== b e)))
\end{verbatim}

\begin{figure}[h!]
  \centering
  \begin{subfigure}{\textwidth}
     \begin{align*}
       \begin{rcases*}
          \{t\ |\ s\} = \{t'\ |\ s'\}\ \land\ C' \\
          \mathtt{tail}(s), \mathtt{tail}(s')\ \text{are not the same variable}
       \end{rcases*} \longmapsto &\ C' \land \textit{any of} \\
       (i)  &\ t = t' \land s = s' \\
       (ii) &\ t = t' \land \{t\ |\ s\} = s'\\
       (iii)&\ t = t' \land s = \{ t' \ |\ s' \} \\
       (iv) &\ s = \{ t'\ |\ N \} \land \{ t\ |\ N \} = s' \land \mathtt{set}(N) \\
     \end{align*}
     \caption{Rule 9 of the $\text{equal}(C)$ solver proposed by Dovier et al. \cite{Dovier2000SetsCLP} }
  \end{subfigure}

  \bigskip

  \begin{subfigure}{\textwidth}
\begin{Verbatim}
(forall (t s t^ s^)
  (forget (== (set-cons t s) (set-cons t^ s^)))
  (scheme (negate eq?) (set-tail s) (set-tail s^))
  =>
  (conde [(== t t^) (== s s^)]
         [(== t t^) (== (set-cons t s) s^)]
         [(== t t^) (== s (set-cons t^ s^))]
         [(fresh (N) (== s (set-cons t^ N)) (== (set-cons t N) s^) (seto N))]))
\end{Verbatim}
  \caption{The above rule translated to \ck{}}
  \end{subfigure}
  \caption{A straightforward translation between existing relational data structure theory and \ck{} syntax}
  \label{fig:rulenine}
\end{figure}

One advantage that moving unification to the constraint level brings is unification over a given semantic domain. For example, consider the finite set solving procedure of Dovier et al.~\cite{Dovier2000SetsCLP}. We can translate the solver rules presented directly into \ck{}, as shown in Figure~\ref{fig:rulenine}. By porting the remaining rules, we translate the complete solver for set unification into \ck{} entirely in userspace, which had previously required a purpose-built interpreter~\cite{mk-beyond-cons}.

\section{Prior Work}
\label{sec:prior}

A large amount of prior work has gone into implementing declarative and extensible constraint solvers, much of which has heavily influenced \ck{}.

\paragraph{CHR(LP)}
Constraint Handling Rules has had a long history of being embedded in logic programming languages: most notably, Prolog~\cite{Schrijvers2005CHR-SWI,Fruhwirth1998CHR}. Much work has been done to make such embeddings efficient, generally by taking advantage of the language's low-level features, such as attributed variables~\cite{LeHuitouze1990} or meta-structures~\cite{Holzbaur1992}. Our work targets \mk{}, a language which is similar to Prolog, but has a number of key differences, especially regarding implementation. \mk{}'s complete search and stream-based semantics are in stark contrast to Prolog's depth-first search and stack-based semantics. To deal with this, we develop a new semantics for \mk{} which incorporates constraint propagation directly.

\paragraph{CLP in \mk{}}
Many implementations of \mk{} ship with some level of constraint solving built-in. These solvers are generally baked directly into the implementation of the language itself. Some implementations of \mk{} --- such as \textsc{cKanren}~\cite{Alvis2011cKanren}, \textsc{constraint-$\mu$Kanren}~\cite{constraintMicroKanren} and \textsc{core.logic}~\cite{corelogic} --- are parameterized over a host-language constraint solver for a particular domain. These solvers must be manually composed, making modularity difficult to achieve. In contrast, \ck{} constraints compose automatically, without the need for host-language solvers or modifications to the language implementation.

\paragraph{\mk{} with Delayed Goals}
One common technique for implementing constraints in userspace is delayed goals, first introduced by Prolog II's \texttt{freeze/2} meta-relation~\cite{Colmerauer1982PrologII}. \textsc{minnaKanren}~\cite{Donahue2023GoalsAsConstraints}, \textsc{Walrus}~\cite{walrus} and the Bachelor's thesis of Zharmukhametova~\cite{microKanrenWithDelayedGoals} provide versions of delayed goal constructors as primitive operators, allowing for some useful user-defined constraints. Delayed goals are, however, limited in their operational potential. They cannot interact with other constraints and so cannot faithfully implement common primitives such as \mk{}'s type constraints due to their disjointedness. \ck{} does not have this limitation, as rules are multi-headed.

\paragraph{Semantics-Guided Program Synthesis (SemGus)}
Semantics-Guided Synthesis~\cite{Kim2021SemGuS} is the task of, for a given language whose semantics is specified as constrained Horn clauses (CHCs), synthesizing a program which satisfies some query. A number of purpose-built solvers for this task exist, such as MESSY~\cite{DAntoni2021Programmable} and ABSYNTHE~\cite{Guria2023Absynthe}, which rely on existing CHC solvers and/or background theory solvers for nontrivial constraint domains such as algebraic data types. \ck{} allows background theories to be specified purely in terms of CHR rules, without the need for an external solver.

\paragraph{FreeCHR}
FreeCHR~\cite{FreeCHR, OpSem4FreeCHR} is a technique which extends CHR to arbitrary host languages by defining a suitable $F$-algebra for evaluation of CHR rules over a given host domain. Such algebras can be composed and interpreted up to a fixpoint, providing a modular way of specifying rules. However, the FreeCHR technique provides no support for partially instantiated variables, unification of user-defined data structures or CHR$^\lor$-style nondeterminism. \ck{} provides these features at the cost of making more assumptions about the host language.

\section{Future Work}
\label{sec:future}

There are several directions in which \ck{} could be extended or improved:

\paragraph{Performance}
\ck{} currently does not take advantage of a number of well-known techniques for making constraint propagation efficient . From the CHR literature, algorithms such as RETE~\cite{rete} and LEAPS~\cite{leaps} provide efficient multi-headed rule matching and constraint compilation, translating CHR rules into efficient host-language code. From the \mk{} literature, multi-staged \mk{}~\cite{pldi-staged-mk} and the optimizing compiler of Ballantyne et al.~\cite{ballantyne2024compiled} provide complementary techniques for improving the performance of search.

\paragraph{Higher-Level Constraint Abstractions}
\phantomsection
\label{par:higherlevel}

In practice, large sets of CHR rules can feel rather ``low-level''. Running rules in all groundedness modes often requires repeating rules for different scenarios. For example, the first two pairs of rules in Listing~\ref{fig:typeo} are semantically identical, but the first of each pair deals with variables which are type-constrained (\verb|(numbero n)| and \verb|(symbolo sym)|, respectively) and the second deals with ground values (\verb|(ground number? n)| and \verb|(ground symbol? sym)|). To our knowledge, all formulations of CHR(LP) currently require this duplication, though we believe a higher-level syntax may make it unnecessary.

\section{Conclusions}
\label{sec:conclusions}

We have presented \ck{}, a dialect of \mk{} that integrates Constraint Handling Rules directly into the language's interpreter, allowing users to write their own constraint solvers declaratively without access to any low-level primitives. We have shown that \ck{}'s constraint propagation mechanism is compatible with \mk{}'s complete, interleaving search strategy, and that the two processes remain complete when composed. We have demonstrated this through two applications: a type-and-example-directed program synthesizer which exhibits finite failure where its purely relational counterpart diverges, and a framework for semantic unification over user-defined data structures.

\section*{Acknowledgments}

We would like to thank Ekaterina Verbitskaia and the members of the \mk{} seminar for an illuminating and interesting discussion on this topic. Will Byrd appreciates the support of Matt Might and the
Hugh Kaul Precision Medicine Institute at the University of Alabama at Birmingham. We would also like to thank our anonymous reviewers for their feedback and references.

\bibliographystyle{eptcs}
\bibliography{my,paper}

\end{document}